\documentclass[submission,copyright]{eptcs}
\providecommand{\event}{ACL2 2011}

\usepackage{breakurl}
\usepackage{color}
\usepackage{amsmath}
\usepackage{graphicx}
\usepackage{amsthm}
\usepackage{amssymb}

\DeclareMathOperator{\escs}{escs}
\DeclareMathOperator{\deps}{deps}

\newcommand{\comment}[1]{}

\begin{document}

\title{Formal verification of a deadlock detection algorithm}
\def\titlerunning{Formal verification of a deadlock detection algorithm}
\author{Freek Verbeek
\institute{Radboud University\\ Nijmegen, The Netherlands}
\institute{Institute for Computing and Information Sciences}
\email{f.verbeek@cs.ru.nl}
\and
Julien Schmaltz 
\institute{Open University \\ Heerlen, The Netherlands}
\institute{School of Computer Science}
\email{julien.schmaltz@ou.nl }
}
\def\authorrunning{Freek Verbeek and Julien Schmaltz}

\maketitle
\begin{abstract}
Deadlock detection is a challenging issue in the analysis and design of on-chip networks. We have designed an algorithm to detect deadlocks automatically in on-chip networks with wormhole switching. The algorithm has been specified and proven correct in ACL2. To enable a top-down proof methodology, some parts of the algorithm have been left unimplemented. For these parts, the ACL2 specification contains constrained functions introduced with defun-sk. We used single-threaded objects to represent the data structures used by the algorithm. In this paper, we present details on the proof of correctness of the algorithm. The process of formal verification was crucial to get the algorithm flawless.
Our ultimate objective is to have an efficient executable, and formally proven correct implementation of the algorithm running in ACL2.
\end{abstract}


\section{Introduction}

Deadlock verification in wormhole networks has been an intricate research area for many years. 
In 1995, Duato proposed a necessary and sufficient condition for deadlock freedom of wormhole networks~\cite{duato95}.
His condition was difficult to understand for many of his peers and required a complex mathematical proof.
In 2010, Taktak et al. were the first to define a polynomial algorithm which can detect deadlocks in wormhole networks automatically~\cite{taktak10}.
In the same year, we formally proved a necessary and sufficient condition of our own~\cite{verbeekschmaltz:tpds10}.
The process of formally proving correctness of this condition helped us recognize a subtle discrepancy in Duato's theorem~\cite{TPDScomment}.
Indeed, due to this discrepancy we could prove that deciding deadlock-freedom in wormhole networks is co-NP-complete, thereby showing Taktak's algorithm had flaws as well.

We have also created an algorithm of our own.
The algorithm has been implemented in C and has achieved good experimental results~\cite{verbeekschmaltz:nocs11}.
Due to the intricacies of deadlock-related theorems in wormhole networks, we wanted a formal proof of correctness to increase our confidence.
To this end, we formalized the algorithm in ACL2\footnote{Proof scripts can be found at \\ http://www.cs.ru.nl/$\sim$freekver/dl\_ic.html}.

Our ultimate objective is to have a formally proven correct and executable algorithm in ACL2.
We want to be able to run this algorithm efficiently on large networks.
To achieve this, we use single-thread objects (stobjs)~\cite{boyer02}.
For now, we have proven correct a \emph{specification} of the algorithm. 
This means that some details have been left unimplemented.
The ACL2 version is not yet executable.
These parts have been replaced by \emph{constrained functions} whose specification is introduced with a {\tt defun-sk} event~\cite{kaufmann01}.
This enables a top-down proving approach.

In this paper we provide some details on the formalization of the algorithm in the ACL2 logic and the proof of correctness.
Due to space limitation, we will not provide much information on the algorithm itself, but focus on the formalized proof of correctness.
For more information, we refer to~\cite{verbeekschmaltz:nocs11}.
Formalizing the algorithm in ACL2 has been of great benefit to us.
The version of the algorithm with which we started had flaws in it, which were detected during the process of theorem proving.

In Section~\ref{sec:wormhole} we shortly introduce wormhole networks and deadlocks.
We explain the basic idea of our algorithm in Section~\ref{sec:algo}.
Section~\ref{sec:formalizing} contains details on formalizing the algorithm in ACL2.
In Section~\ref{sec:proof} we provide details on the proof of correctness.
We conclude in Section~\ref{sec:conclusion}.

\section{Wormhole networks}\label{sec:wormhole}

In wormhole networks, messages are decomposed into data units called \emph{flits}. A flit constitutes the atomic object 
that is transferred between any two channels. Typically, there is a header flit followed by 
a sequel of data flits. The end of a packet is marked by a tail flit. For simplicity, we do 
not distinguish between data flits and the tail flit. We refer to all of them as the tail. Only the header flit contains information on the destination of the message. 
The header flit advances along the specified route, while the tail follows in a pipe-line 
fashion. When the header flit is blocked, all flits of the message are blocked. A channel 
can only store flits belonging to at most one message. Therefore, tail flits block header 
flits of other messages.

In~\cite{verbeekschmaltz:tpds10} we have proven a necessary and sufficient condition for deadlock-free routing in wormhole networks. This proof has been formalized in ACL2. We shortly address this condition. In wormhole networks, messages occupy paths of channels in the network. A path that can be occupied by a message destined for $d$ will be called a \emph{$d$-path}. As flits in the tail follow the header flit, blockage of a message depends solely on the header flit. The central idea of the condition is that a header flit must always have an \emph{escape}. An escape is a next hop supplied by the routing function for the destination of the message. The escape must be available, i.e., not occupied by other worms.

\begin{figure}[hbpt]
\centering
\includegraphics[scale=0.25]{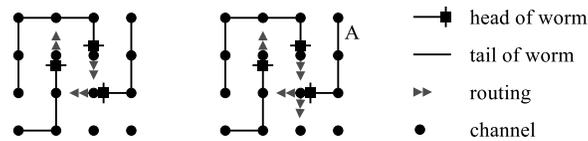}
\caption{Wormhole configurations}
\label{fig:deadlock}
\end{figure}

Consider the first configuration in Figure~\ref{fig:deadlock}. Three messages occupy three paths of channels. For each path, the head of the path cannot escape as the routing function does not supply next hops that are not included in the set of paths. There exists a set of paths without an escape, which corresponds to the existence of a deadlock. In the second configuration, the header flit of message $A$ is supplied two possible next hops for its destination. As one of them is not included in the set of paths, the header flit can move towards this escape and resolve the deadlock. The set of next hops has an escape and is therefore not a deadlock.
Our condition states that:
\begin{center}
A wormhole network is deadlock-free if and only if for any pairwise disjoint set of $d$-paths there exists an escape.
\end{center}

Checking this condition is co-NP-complete~\cite{verbeekschmaltz:tpds10}. Our algorithm is polynomial, but may return a false deadlock. It returns a set of paths without an escape if there exists such a set or returns {\tt t} if there exists no such set. The set of paths is however not necessarily pairwise disjoint.

\section{Algorithm}\label{sec:algo}

The basic objective of our algorithm is to mark each channel.
After termination of the algorithm, either all channels are marked to be immune for deadlock, or a deadlock can be constructed from those channels that have not been marked immune for deadlock.
We use the following markings:
\begin{description}
\item[$\mathbf{0}$] The channel is unmarked.
\item[$\mathbf{1}$] The channel has been visited, but a definite mark has not yet been determined. 
\item[$\mathbf{2}$] The channel is immune for deadlock, i.e., no flit in the channel can be permanently blocked.
\item[$\mathbf{3}$] There exists a destination $d$ such that a header flit destined for $d$ can be permanently blocked.
\item[$\mathbf{4}$] No header flit can be permanently blocked, but for some destination $d$ a tail flit can be permanently blocked.
\end{description}
After termination, all channels are marked either $\mathbf{2}$, $\mathbf{3}$, or $\mathbf{4}$. If all channels are marked $\mathbf{2}$, then the network is deadlock-free. A $\mathbf{2}$-marked channel is always immune for deadlock. If channels have been marked either $\mathbf{3}$ or $\mathbf{4}$, a deadlock can be constructed. A $\mathbf{3}$-marked channel $c$ can be filled with a header flit destined for $d$.
A $\mathbf{4}$-marked channel $c$ can be filled with tail flits. 

The algorithm obtains these markings by checking for each channel $c$ and for each destination $d$ the possible next hops.
If for some destination $d$ there is no next hop marked $\mathbf{2}$, then a header flit with destination $d$ can be permanently blocked, as all next hops can be permanently blocked.
The channel is marked $\mathbf{3}$.
If for channel $c$ for all destinations there exists a $\mathbf{2}$-marked neighbor, then channel $c$ cannot be marked $\mathbf{3}$.
If in this case there exists a $d$-path leading to $\mathbf{3}$-marked channel $h$, this path can be filled with tail flits. As in channel $h$ a header flit can be permanently blocked, the tail flits in channel $c$ can be permanently blocked. Channel $c$ is marked $\mathbf{4}$. Otherwise the channel is marked $\mathbf{2}$, as it is immune for deadlock.

\begin{figure}[htbp]
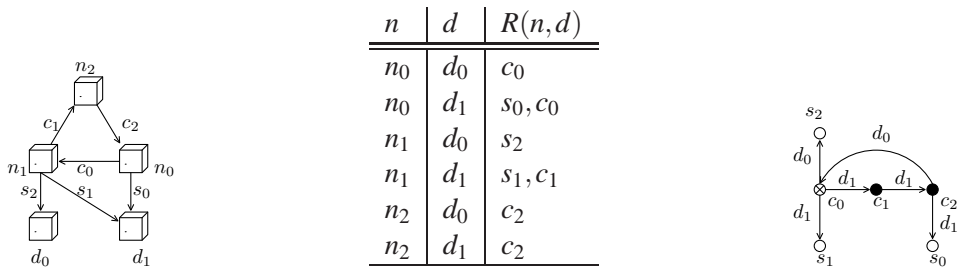

\centering
\begin{minipage}[b]{0.32\linewidth}\centering
        \includegraphics[scale=0.15]{pics/network.1}
\end{minipage}
\begin{minipage}[b]{0.32\linewidth}\centering
        \begin{tabular}{l|l|l}
        $n$ & $d$ & $R(n, d)$\\\hline\hline
        $n_0$ & $d_0$ & $c_0$\\
        $n_0$ & $d_1$ & $s_0, c_0$\\
        $n_1$ & $d_0$ & $s_2$\\
        $n_1$ & $d_1$ & $s_1, c_1$\\
        $n_2$ & $d_0$ & $c_2$\\
        $n_2$ & $d_1$ & $c_2$\\\hline
        \end{tabular}

        \vspace{0em}
\end{minipage}
\begin{minipage}[b]{0.32\linewidth}\centering
        \includegraphics[scale=0.15]{pics/dep-graph.1}
\end{minipage}
\caption{Example network, routing and graph}
\label{fig:ex-trace}
\end{figure}

Consider the network in Figure~\ref{fig:ex-trace}. In the network, messages generated in the processing nodes $n_0$ to $n_2$ move from channel to channel. Nodes $d_0$ and $d_1$ are the only possible destinations. Figure~\ref{fig:ex-trace} also shows the routing function.
The graph representation of the network is the input of the algorithm.
An edge $(c_0, c_1)$ is labelled $d$ if a message in channel $c_0$ destined for $d$ is routed towards $c_1$.

Destinations $d_0$ and $d_1$ are sinks. They can never be blocked. Channels $s_0$, $s_1$, and $s_2$ are marked with $\mathbf{2}$ as they are immune for deadlock.
Channel $c_2$ is marked $\mathbf{3}$, as for destination $d_0$ all neighbors are not marked $\mathbf{2}$.
Similarly, channel $c_1$ is marked $\mathbf{3}$.
Lastly, channel $c_0$ is marked $\mathbf{4}$. For all destinations, there is $\mathbf{2}$-marked neighbor, but there exists a path leading to $c_1$ which is marked $\mathbf{3}$.

There exists exactly one possible deadlock-configuration: channels $c_0$ and $c_1$ are filled with a worm with destination $d_1$ and channel $c_2$ is filled with a header flit destined for $d_0$.
The deadlock can be obtained by filling $\mathbf{3}$-marked channels with header flits and $\mathbf{4}$-marked channels with tail flits. 

\section{Formalization in ACL2}\label{sec:formalizing}

\begin{figure}[hbpt]
\centering
\includegraphics[scale=0.1]{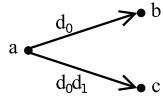}
\caption{Routing represented in a graph}
\label{fig:graph}
\end{figure}

First, we define the data structure in which the graph is stored. The graph consists of vertices $0, 1, \ldots, C-1$, with $C$ the number of channels in the network. With each channel $c$ a list of neighbors is associated, representing the possible next hops a message in channel $c$ can take. The labels on the edges represent the destinations which cause a message to be routed towards the neighbors. For example, the graph in Figure~\ref{fig:graph} represents a network where a message in channel $a$ can be routed towards channel $b$ for destination $d_0$ and to channel $c$ for destinations $d_0$ and $d_1$. 

In ACL2, we store this data structure in a stobj {\tt graph}. Function {\tt (neighbors c d graph)} takes as parameters a channel $c$, a destination $d$, and the stobj {\tt graph} and returns a list of neighbors. For sake of clarity, we will not mention this stobj any further.

The algorithms needs to store markings. These are stored in a stobj {\tt marks}. For each channel $c$, we store a marking between $0$ and $4$, a list $\escs(c)$ of destinations on edges leading to $\mathbf{2}$-marked neighbors (escapes) and a list $\deps(c)$ of destinations on edges not leading to $\mathbf{2}$-marked neighbors. All channels are initially unmarked.
\begin{verbatim}
(defstobj marks
  (marks :type (array (integer 0 4) (C))
                                :initially 0)
  (escs  :type (array list (C)) :initially nil)
  (deps  :type (array list (C)) :initially nil)
\end{verbatim}
ACL2 introduces functions to access this stobj. For example, to obtain the marking of channel $c$, we can use:
\begin{center}
{\tt (marksi c marks)}
\end{center}

Formalizing the algorithm in ACL2 was a straightforward exercise in LISP. For now, we have left some parts of the algorithm unimplemented. Because of this, we were able to prove correctness of the algorithm, regardless of how these parts are implemented. Also, this approach enabled a top-down proving approach, as we could first prove correctness of the algorithm as a whole without getting stranded in the details. An example is that at some point the algorithm marks a channel $c$ with $\mathbf{4}$ if there exists a path that satisfies some properties. The path must be traversable by a message destined for some destination $d$ (i.e., it must be a $d$-path), start in $c$, end in a channel $c_e$ that is not marked $\mathbf{2}$, and destination $d$ must have been added by the algorithm to $\deps(c_e)$ but not to $\escs(c_e)$. An efficient decision procedure for the existence of such a path is an algorithm of its own. At this point, we do not want to bother ourselves with this, as it is only a small part of the algorithm. We therefore replace this decision procedure with an unimplemented specification, introduced by a {\tt defun-sk} construct:
\begin{verbatim}
(defun-sk ex-d-path-to-not2(c marks)
 (exists (p d)
  (let ((start (car p))
        (end (car (last p))))
   (and (d-pathp p d)
        (equal start c)
        (not (equal (marksi end marks) 2)
        (member-equal d (depi end marks))
        (not (member-equal d (esci end marks)))
)))))
\end{verbatim}
Function {\tt d-pathp} is a recognizer for paths that can be established by the routing function for destination $d$.

The algorithm calls function {\tt ex-d-path-to-not2}. At a later stage, an implementation can be made and proven correct with respect to this specification.

\section{Proving correctness}\label{sec:proof}

The proof of correctness consists of two parts: if the algorithm returns {\tt t} there is no set of paths without an escape and if the algorithm returns {\tt nil} there is a set of paths without an escape. In this paper, we give details on the second part of the proof, i.e., we show that from the markings a set of paths without an escape can be constructed.

\subsection{Informal proof of correctness}

\begin{proof}
If the algorithm marks a channel $\mathbf{3}$ or $\mathbf{4}$, it is possible to create a set of paths without an escape. This proof formalizes the intuition in Figure \ref{fig:ex-trace}: a deadlock is created from all $\mathbf{3}$- and $\mathbf{4}$-marked channels. 
\begin{enumerate}
\item Take the set of paths $\Pi_\mathbf{34}$ obtained by taking -- after termination -- for each $\mathbf{3}$-marked channel $c$ the singleton path $[c]$ and for each $\mathbf{4}$-marked channel a path leading to a $\mathbf{3}$-marked channel.
\item Each $\mathbf{3}$-marked channel $c$ in the set of paths $\Pi_\mathbf{34}$ has a destination $d$ that is a member of $\deps(c)$ and not of $\escs(c)$, since channels are marked $\mathbf{3}$ only if $\deps(c) \nsubseteq \escs(c)$.
\item Since, if some destination leads to $\mathbf{2}$-marked neighbors it is added to $\escs(c)$, destination $d$ does not lead to neighbors marked $\mathbf{2}$.
\item Since destination $d$ does not lead to $\mathbf{2}$-marked neighbors, it leads to channels marked $\mathbf{3}$ or $\mathbf{4}$ only.
\item Since the set of paths $\Pi_\mathbf{34}$ contains all $\mathbf{3}$- and $\mathbf{4}$-marked channels and since channel $c$ has destination $d$ which leads to $\mathbf{3}$- and $\mathbf{4}$-marked channels only, channel $c$ is not an escape for this set of paths (i.e., all its neighbors for destination $d$ are included in the set of paths). Thus $\mathbf{3}$-marked channels are no escapes.
\item As for all $\mathbf{4}$-marked channels there exists a path leading to a $\mathbf{3}$-marked channel, these are no escape either.
\item Since none of the channels in the set of paths $\Pi_\mathbf{34}$ is an escape, the set of paths has no escape.
\item The algorithm returns true if and only if after termination there exists at least one $\mathbf{3}$- or $\mathbf{4}$-marked channel. Thus it returns true if there exists a non-empty set of paths without an escape.
\end{enumerate}
\end{proof}

\subsection{Formal proof of correctness}

We provide some details on formalizing the informal proof. We will not consider all steps, but focus on some of the interesting aspects.

\subsubsection{Step 1: constructing a witness}
In this step we need to construct a witness $\Pi_\mathbf{34}$ of which we are going to prove that it is a set of paths without an escape. In Step 1, it is implicitly assumed that for all $\mathbf{4}$-marked channels
 there actually exists a path leading to a $\mathbf{3}$-marked channel. We first express this assumption using a {\tt defun-sk} construct.
\begin{verbatim}
(defun-sk ex-d-path-to-3(c marks)
 (exists (p d)
  (and (d-pathp p d)
       (equal (car p) c)
       (equal (marksi (car (last p)) marks) 3))))
\end{verbatim}
For some destination $d$ there exists a $d$-path $p$ starting in the given channel $c$ and ending in a $\mathbf{3}$-marked channel. 

Now we build the witness, i.e., a set of paths, using the witness introduced by the {\tt defun-sk}:
\begin{verbatim}
(defun witness-set-of-paths (n marks)
 (declare (xargs :non-executable t))
 (cond
  ((zp n) nil)
  ((equal (marksi (1- n) marks) 3)
   (cons (list (1- n))
         (witness-set-of-paths (1- n) marks)))
  ((equal (marksi (1- n) marks) 4)
   (cons
    (car (ex-d-path-to-3-witness (1- n) marks))
    (witness-set-of-paths (1- n) marks)))
  (t
   (witness-set-of-paths (1- n) marks))))
\end{verbatim}
For each $\mathbf{3}$-marked channel a singleton path {\tt (list c)} is created. For each $\mathbf{4}$-marked channel, the witness introduced by the {\tt defun-sk} construct is used.
Here we run into a problem: {\tt marks} is a stobj storing the markings. However, a defun-sk cannot declare parameters to be stobjs. If we would add the declaration
\begin{verbatim}
(declare (xargs :stobjs (marks)))
\end{verbatim}
to function {\tt witness-set-of-paths}, as we ordinarily would want to do, ACL2 produces an error that a single-threaded object, namely {\tt marks}, is being used where an ordinary object is expected.
 Our solution was to omit this declaration, meaning that {\tt marks} is not considered a stobj, but can be any ordinary object. However, this means that we cannot use the standard accessor function {\tt marksi} to access the stobj {\tt marks}, as {\tt marks} is not declared to be the stobj {\tt marks}. If we declare the function to be non-executable, this problem is solved. We have a function generating a witness, it is however not executable.

Now we need to prove that after termination, for each $\mathbf{4}$-marked channel there exists a path leading to a $\mathbf{3}$-marked channel, i.e., we need to establish that {\tt (ex-d-path-to-3 c marks)} holds for all $\mathbf{4}$-marked channels $c$. This is an inductive invariant.
We express the invariant:
\begin{verbatim}
(defun invariant-4marks (n marks)
 (declare (xargs :non-executable t)) 
 (cond
  ((zp n) t)
  ((equal (marksi (1- n) marks) 4)
   (and (ex-d-path-to-3mark (1- n) marks)
        (invariant-4marks(1- n) marks)))
  (t
   (invariant-4marks (1- n) marks))))
\end{verbatim}
We need to prove that each line of code of the algorithm preserves this invariant under some assumptions. As an example, the following theorem expresses that marking a channel $\mathbf{2}$ preserves the invariant:
\begin{verbatim}
(defthm mark2-preserves-invariant-4marks
 (let ((marks-after (update-marksi c 2 marks))
  (implies (and (invariant-4marks n marks)
                (not (equal (marksi c marks) 3)))
    (invariant-4marks n marks-after)))))
\end{verbatim}
If a channel $c$ is marked $\mathbf{2}$ and it was not marked $\mathbf{3}$, the invariant is preserved. This holds, since the witness $\pi$ before setting the $\mathbf{2}$-mark is also a witness after setting the mark. 
For each line of code of the algorithm, a theorem similar to this has been proven. We also need to prove that initially the invariant holds:
\begin{verbatim}
(defthm forall-unmarked-implies-invariant-4marks
  (implies (forall-unmarked n marks)
           (invariant-4marks n marks)))
\end{verbatim}
Function {\tt forall-clear} expresses that all markings are clear, i.e., they are all set to $\mathbf{0}$. The proof of this theorem is trivial, as there are no $\mathbf{4}$-marked channels.

\subsubsection{Step 2: more invariants}
Step 2 is basically just an invariant. 
\begin{verbatim}
(defun invariant-3marks (n marks)
 (cond
  ((zp n) t)
  ((equal (marksi (1- n) marks) 3)
   (and (not (subsetp (depi (1- n) marks)
                      (esci (1- n) marks)))
        (invariant-3marks (1- n) marks)))
 (t
  (invariant-3marks (1- n) marks))))
\end{verbatim}
For each $\mathbf{3}$-marked channel $c$, there exists a destination in $\deps(c)$ that is not in $\escs(c)$.
The proof proceeds similar to the proof of the invariant used in Step 1. For each line of code of the algorithm, a theorem is proven that the line preserves the invariant.

The same methodology applies to Steps 3 and 4 of the informal proof. This introduces more invariants on each marking.

\subsubsection{Step 5: correctness of witness}
At this point, we have established correctness of the invariants and proven them inductive. Now we use the invariants to prove theorems on the constructed witness. For example, we prove in step 5 that a $\mathbf{3}$-marked channel is not an escape for the set of paths $\Pi_\mathbf{34}$.
\begin{verbatim}
(defthm r-marked-3-->no-escape-for-witness
 (let ((d (find-member-not-in (depi c marks)
                              (esci c marks))))
  (implies
   (and (equal (marksi c marks) 3)
        (invariant-3marks C marks)
\end{verbatim}
\vspace{-1em}\hspace{10ex}$\vdots$\hspace{2ex}{\tt invariants}
\vspace{-1em}
\begin{verbatim}
        )
   (subsetp
    (neighbors c d)  
    (union-of (witness-set-of-paths C marks))))))
\end{verbatim}
Function {\tt find-member-not-in} takes two lists and returns an element from the first list that is not in the second. We use it to find the destination $d$ that is in $\deps(c)$ but not in $\escs(c)$. Assuming all the invariants needed to prove this theorem, we prove that the set of neighbors of $c$ for destination $d$ is a subset of the union of the set of paths $\Pi_\mathbf{34}$.
It is therefore not an escape for this set of paths.

The proof of Step 6 is done in a similar fashion. Step 7 follows by definition.

\subsubsection{Step 8: final theorem}
\begin{figure*}[ht]
\centering
\begin{minipage}{0.75\textwidth}
\begin{verbatim}
(defthm algo-returns-nil-->deadlock
 (let ((marks-after-termination (mv-nth 1 (algorithm marks)))
       (p-witness (witness-set-of-paths C marks-after-termination))
       (d-witness (witness-set-of-dests C marks-after-termination))
       (l-witness (len p-witness))))
  (implies (and (forall-clear C marks)
                (equal (mv-nth 0 (algorithm marks)) nil))
           (and (> l-witness 0)
                (set-of-paths-witnessp l-witness p-witness d-witness))))
\end{verbatim}
\end{minipage}
\caption{Final theorem}
\label{fig:finaltheorem}
\end{figure*}

The final theorem that we prove in this paper states that if our algorithm returns {\tt nil}, there exists a set of paths without an escape.

We first define a recognizer for such sets of paths:
\begin{verbatim}
(defun set-of-paths-witnessp (n paths dests)
 (if (zp n)
  (and (endp paths) (endp dests))
  (let ((p (nth (1- n) paths))
        (d (nth (1- n) dests))
   (and (subsetp (neighbors (car(last p)) d)
                 (union-of paths))
        (d-pathp p d)
        (set-of-paths-witnessp (1- n) paths dests)
)))))
\end{verbatim}
The function takes as input a list of paths and a list of the destinations for which these paths are established. Also, it takes as input the number of paths. It checks if for each $d$-path $p$ the neighbors of the last channel (where the head of the worm is located) cannot escape the paths.

Figure~\ref{fig:finaltheorem} gives the final theorem.
The algorithm returns a multi value with as first value a boolean $b$ which is {\tt t} if and only if there is no deadlock. The second value is the stobj {\tt marks} after termination. We have a function generating the witness for the destinations of the paths, similar to the function generating the witness for the paths themselves (see Step 1). If initially all markings are clear, all inductive invariants can be proven and theorems such as the theorem in Step 5 can be applied to prove correctness of the witnesses.  We also prove that the witness is non-empty.

\section{Conclusion}\label{sec:conclusion}

We have formally proven correctness of an algorithm which detects deadlocks in wormhole networks.
The process of theorem proving has been crucial for us to get all the details right.

The entire proof consists of 7263 lines of ACL2 code. A great part of this consists of proving that each line of the algorithm preserves each of the invariants.
Proving correctness of the invariants was a relatively easy process.
The theorem to be proved is similar each time: there is an invariant which holds initially, and it must hold -- under some assumptions -- after executing one line of code.
The trick is to find these assumptions, but these can be figured out from the output of ACL2.

The use of {\tt defun-sk} allowed us to leave some parts of the algorithm unimplemented and replace them with a specification of what the code should do.
This allowed us to start with a global proof before stranding into details.
As we could proceed with the proof, we could first see whether the specification was correct before making an implementation.
If the specification was insufficient to finish the proof, we could simply change the specification and did not have to reimplement some part of the algorithm.

The algorithm is not yet executable, as some parts have been left unimplemented. Future work consists of implementing these parts efficiently and proving them correct with respect to the specification that is currently used. 
Once the algorithm is executable, we can compare the performance to our C implementation.
Our ultimate objective is to have a fully formally verified and efficiently executable implementation in ACL2.

\nocite{*}
\bibliographystyle{eptcs}
\bibliography{ref.bib}

\end{document}